\documentclass[11pt,a4paper]{article}
\usepackage{setspace}
\usepackage{epsfig}
\begin{document}

\title{\bf\large The process $e^+e^-\to\pi^+\pi^-\pi^0$
in the energy range $2E_0=1.04$--1.38~GeV}
\author{\normalsize M.N.Achasov, V.M.Aulchenko, S.E.Baru, K.I.Beloborodov,
A.V.Berdyugin,\\ \normalsize A.V.Bozhenok, A.D.Bukin, D.A.Bukin, S.V.Burdin,
T.V.Dimova, S.I.Dolinsky,\\ \normalsize V.P.Druzhinin, M.S.Dubrovin,
I.A.Gaponenko, V.B.Golubev, V.N.Ivanchenko, \\ \normalsize I.A.Koop, P.M.Ivanov,
A.A.Korol, S.V.Koshuba, E.V.Pakhtusova,\\ \normalsize E.A.Perevedentsev,
A.A.Salnikov, 
\underline{S.I.Serednyakov\thanks{e-mail:serednyakov@inp.nsk.su; FAX:
+7(3832)342163}},
V.V.Shary, Yu.M.Shatunov,\\ \normalsize
V.A.Sidorov, Z.K.Silagadze, A.N.Skrinsky, Y.V.Usov, A.V.Varganov,\\ 
\normalsize A.V.Vasilyev, Yu.S.Velikzhanin\\ \\
\normalsize Budker Institute of Nuclear Physics, Novosibirsk State University,\\ 
\normalsize Novosibirsk, 630090, Russia}
\date{}
\maketitle

\begin{abstract}
\normalsize
In the experiment with the SND detector at VEPP-2M $e^+e^-$ collider
the process $e^+e^-\to\pi^+\pi^-\pi^0$ was studied in the
energy range $2E_0$ from 1.04 to 1.38~GeV. A broad peak was observed with
the visible mass $M_{vis}=1220\pm 20$ MeV and cross section in the 
maximum $\sigma_0\simeq 4$ nb. The peak can be interpreted as a 
$\omega$-like resonance  $\omega (1200)$.  
\end{abstract}

\twocolumn

{\large\bf Introduction.}
\normalsize The process  $e^+e^-\to\pi^+\pi^-\pi^0$ dominates in the isoscalar 
part of the total cross section of  $e^+e^-$ annihilation into hadrons.
In the low energy  region $2E_0\sim$1~GeV this process
is measured with a relatively high 
accuracy $\sim 3\%$ only in the vicinity of isoscalar
resonances $\omega$(783) and $\phi$(1020). In the region above  $\phi$(1020)
there are data from detectors ND \cite{ND} and DM2 \cite{DM2}, but the
statistics in these experiments is quite small. In the Tables \cite{PDG} the
$\omega$(1420) and $\omega$(1600) states are listed, 
but their parameters, based entirely
on DM2 measurements of the processes $e^+e^-\to 3\pi ,\omega\pi\pi$,
are not well established.

{\large\bf Experiment.} 
\normalsize  In this work the process
            \begin{equation} \label{3pi}
              e^+e^-\to\pi^+\pi^-\pi^0
             \end{equation}
was studied in the energy range $2E_0=1.04-1.38$~GeV at the VEPP-2M
$e^+e^-$ collider with SND detector \cite{SND}. The integrated luminosity in the
experiment \cite{MHAD} is $L=6.1\mathrm{pb}^{-1}$.
In the study of the process  (\ref{3pi})
the main background comes from the processes
              \begin{equation} \label{3pi,gamma}
              e^+e^-\to\pi^+\pi^-\pi^0\gamma,
	      \end{equation}

              \begin{equation} \label{4pi}
	      e^+e^-\to\pi^+\pi^-\pi^0\pi^0,
	      \end{equation}

              \begin{equation} \label{QED}
	      e^+e^-\to e^+ e^-\gamma\gamma.
	      \end{equation}

  The radiative process  (\ref{3pi,gamma}) with emission of hard photons
particularly by initial electrons is a source of  background in the vicinity
of $\phi (1020)$, where its cross section is determined by intermediate 
$\phi\gamma$ state decaying further into $\pi^+\pi^-\pi^0\gamma$.
The process (\ref{4pi}) gives contribution due to 
merging of showers in the calorimeter, loss of photons and errors
in events reconstruction.
              
{\large\bf Events Selection.}  To select events of the process (\ref{3pi}) and
suppress the background, the following cuts  were applied:
\begin{enumerate}
\item Two charged particle tracks and two photons are found in an event,
\item Particle angle with respect to the beam  is $\theta >27^o$,
\item The total energy deposition $\Delta E$ in 
 is $0.9E_0<\Delta E<1.8E_0$,
\item The spatial angle $\psi_{12}$ between charged particles 
is $\psi_{12}<160^o$,
\item The reconstructed by kinematic fitting 
energies of two charged  pions are restricted  by the cuts
$E_1<0.75\cdot E_0$ and $E_2<0.65\cdot E_0$,
\item The minimal photon energy is $E_{\gamma ,min}>0.1\cdot E_0$,
\item The photon quality parameter  is $\zeta<30$,
\item The kinematic fit parameter is $\chi_{3\pi}<10$.
\end{enumerate}

   The photon quality parameter  $\zeta$ describes the likelihood of
a hypothesis, that given transverse energy profile of a cluster in the
calorimeter can be attributed to a single photon \cite{phot}.
The parameter  $\chi_{3\pi}$ describes the degree  of energy-momentum
balance in an event under assumption of $\pi^+\pi^-\pi^0$ final state.
The cut 8 suppresses all background  processes, while other cuts 
are efficient against the processes  (\ref{QED}) and  (\ref{4pi}).

\begin{figure}[htb]
   \epsfig{figure=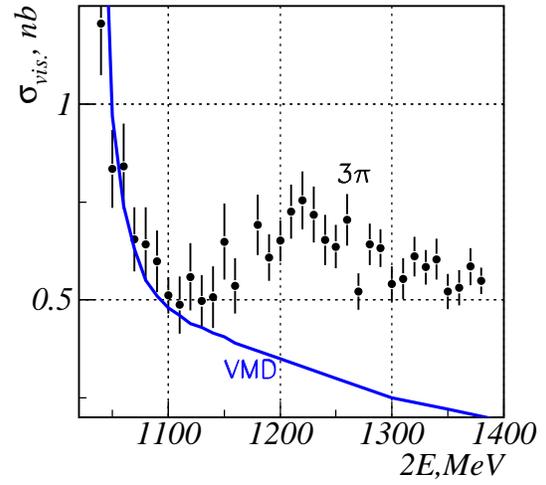,width=0.45\textwidth}
   \caption{ \label{vis}  The visible cross section of the process
   $ e^+e^-\to\pi^+\pi^-\pi^0$. The solid line shows the prediction
   of vector meson dominance model (VMD)
   for the processes (\ref{3pi}) and  (\ref{3pi,gamma}).}
   \end{figure}

{\large\bf Data analysis.}  The total of N=6550 events were selected in
34 energy points after applying selection cuts.
The visible cross section $\sigma_{vis}=N/L-\sigma_B$, obtained after imposing all
selection cuts, is shown in the Fig.\ref{vis}. 
In the definition of $\sigma_{vis}$, $L$ is an integrated luminosity 
in a given energy point, measured by means of
Bhabha scattering process, $N$ is the number of selected events, and
$\sigma_B$ is the contribution from the background processes
(\ref{4pi}) and  (\ref{QED}). The value of $\sigma_B$, estimated from
simulation,  does not exceed 5\%. 
The solid line in Fig.\ref{vis}
shows the vector meson dominance model (VMD) prediction
for the processes (\ref{3pi}) and  (\ref{3pi,gamma}) with contributions
of $\omega (783)$ and $\phi (1020)$ states only. One can see,
that at $2E_0>1100$~MeV, the measured cross section significantly
exceeds VMD prediction. An enhancement in the visible cross
section near 1200~MeV is seen.

\begin{figure}[htb]
   \epsfig{figure=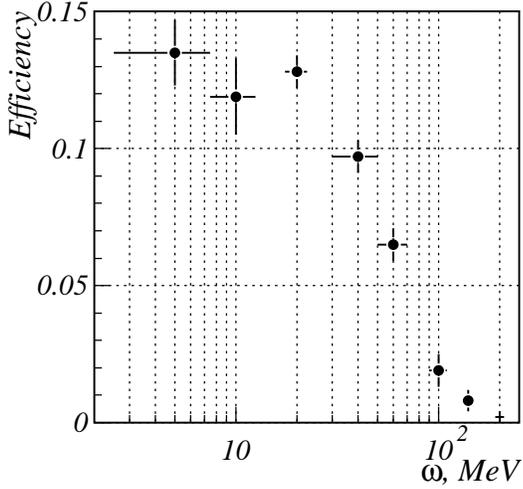,width=0.45\textwidth}
   \caption{ \label{eff} The  detection efficiency  of the process   (\ref{3pi,gamma}) 
versus radiative photon energy $\omega$.    At the low photon energy  $\omega <$10 MeV
the value of the detection efficiency is the same as for the process  (\ref{3pi})}.
   \end{figure}

          The detection efficiency $\epsilon$ for the process  (\ref{3pi})
was calculated using simulated events of the processes (\ref{3pi}) and 
(\ref{3pi,gamma}).
The obtained value of $\epsilon$ varies from 12\%  to  14\%  
in the energy  range 2E from 1040 to  1380 MeV.
It was found, that  for the process  (\ref{3pi,gamma}) the detection
efficiency  strongly
depends on the energy of the radiative photon $\omega$ and
due to the simple relation: $\omega\simeq 2E_0-M_{\phi}$ ---
on the beam energy. One could see from  Fig.\ref{eff}, that  events  with
$\omega > 100$~MeV do not pass the cuts. Thus at the beam energy
$2E > 1120$~MeV the
$\phi\gamma$ intermediate state in the process (\ref{3pi,gamma})
does not contribute into  $\sigma_{vis}$ and the detection efficiency
is determined mainly by the process (\ref{3pi}).

\begin{figure}[htb]
   \epsfig {figure=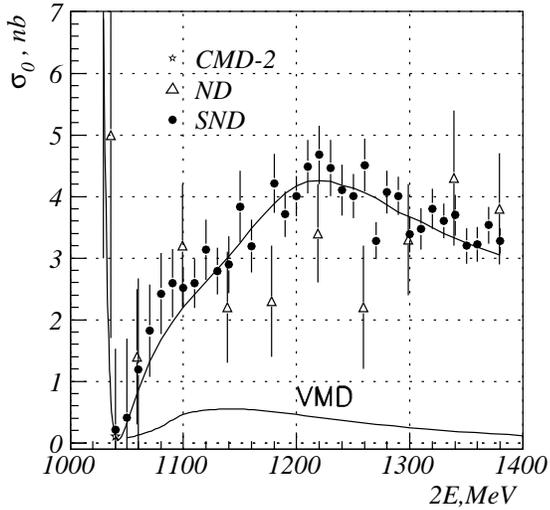,width=0.45\textwidth}
   \caption{ \label{p3} The $e^+e^-\to\pi^+\pi^-\pi^0$ total cross section,
   measured by VEPP-2M detectors ND \cite{ND}, CMD-2 \cite{CMDP3},
   and SND (this work). 
The only   CMD-2 point at 2E=1040 MeV with the cross section of 0.1 nb 
is not clearly seen,  because it  overlapes  with the nearest SND point.
  The upper solid line is the result of fitting of
   all existing experimental data according to Eq.\ref{fit}.
   The lower curve is a VMD model prediction.}
   \end{figure}

\begin{figure}[htb]
    \epsfig{figure=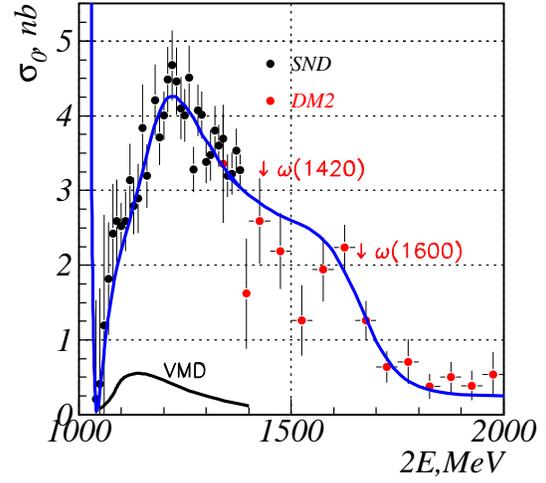,width=0.45\textwidth}
   \caption{\label{sdm2} The cross section of the process
    $e^+e^-\to\pi^+\pi^-\pi^0$  from this work  and  DM2 \cite{DM2}
    experiment. The upper solid line is the  fit curve, obtained with 
    Eq.\ref{fit}.  The lower curve is a VMD model prediction.}
   \end{figure}

   To determine the $\pi^+\pi^-\pi^0$ production cross section the
following expression was used:
        \begin{equation} \label{Appr}
\sigma_{vis} = \epsilon\cdot\sigma_0\cdot (1+\delta ),
	\end{equation}
where   $\sigma_0$ is the cross section of the process (\ref{3pi}), $\delta$ is a
radiative correction \cite{rad}.
The radiative correction sharply depends on the energy, decreasing from $\sim$50
at $2E_0=1040$~MeV to  $\sim$0.2 at  $2E_0=1200$~MeV. All variables in Eq.\ref{Appr} 
are considered as functions of energy.
The  Born cross section  $\sigma_0$ of the process (\ref{3pi}),
obtained from the
Eq.\ref{Appr} is shown in Fig.\ref{p3} and listed in the Table \ref{tabl1}.
The broad peak is seen with effective mass $M_{eff}\simeq 1200$~MeV.
One can see, that  the  measured cross section agrees with
the previous ND data \cite{ND} and recent CMD-2 measurements
at $2E<1050$~MeV \cite{CMDP3}.
In simulation of the processes (\ref{3pi}) and (\ref{3pi,gamma}) 
the cross section energy dependence was taken from the Table \ref{tabl1}.
The measured cross section $\sigma_0$ is plotted in a wider energy range   in the
Fig.\ref{sdm2} together with DM2 data \cite{DM2}. One can see, that DM2 cross
section at  higher energy well matches SND data.

   The systematic error of the measured cross section
includes following contributions:
\begin{itemize}
\item[-]{error in detection efficiency estimation $\sim$10\%,}
 \item[-]{error in background subtraction from
the processes (\ref{4pi}), (\ref{QED})  $\sim$3\%,}
\item[-]{error in
background subtraction from the radiative process (\ref{3pi,gamma}).}
\end{itemize} 
  The last error is estimated to be  $\sim$3\%
at $2E_0>1150$~MeV, but at lower energy it
grows up to  $\sim$50\%. The total systematic error at
$2E_0>1150$~MeV is estimated to be 12\%.

 The structure of the $3\pi$ final state was analysed  in our earlier work
\cite{MHAD} where it was found, that $\rho\pi$ intermediate state
dominates there. 
Moreover, manifestation of $\rho -\omega$ interference in the
final $3\pi$ state, caused by $e^-e^+\to\omega\pi^0$, $\omega\to\pi^-\pi^+$
process, predicted in \cite{romeg}, was observed. The immediate
consequence of this effect is possible change in
mass spectra and cross section of the process  (\ref{3pi})
by $\sim$10\%. In the present work  the process (\ref{3pi}) 
was simulated with the  only $\rho\pi$  intermediate state.

{\large\bf Fitting of the cross section.}
To test energy dependence of the cross section on possible deviations
from VMD model, we fitted our data together with the data 
obtained in other experiments outside the interval $2E=1.04\div 1.38$ GeV: 
 ND\cite{ND}, CMD-2\cite{CMD2}, and DM2\cite{DM2}.

The following expression  from the works \cite{Wrhopi,CMDP3} was 
used for approximation of the cross section as a sum of resonances:
   \begin{equation} \label{fit}
   \begin{array}{l}
 \sigma_0 (e^+e^-\to \pi^+\pi^-\pi^0)=\frac{\displaystyle W_{\rho\pi}(s)}
 {\displaystyle s^{3/2}}\cdot \\ \left| \sum\limits_{V} \sqrt{\frac
 {\displaystyle  \sigma_V\cdot m^3_V}{\displaystyle W_{\rho\pi}
(m_V^2)}}\cdot\frac{\displaystyle e^{i\phi_V}\Gamma_Vm_V} 
{\displaystyle s-m^2_V-im_V\Gamma_V(s)}\right|^2,\\
\sigma_V=\frac{\displaystyle 12\pi B_{Vee}B_{V\rho\pi}}{
\displaystyle m^2_V}.
   \end{array}
   \end{equation}
Here $W_{\rho\pi}(s)$ is a phase space factor of the final state.
The following 4 resonances were included in the fitting:  $\omega (783)$,
$\phi (1020)$, $\omega (1600)$, and an additional $\omega$-like state
 $\omega (1200)$ with its parameters set free. The energy dependence 
of $\Gamma_V$ was taken into account only for two lightest
 states  $\omega (783)$ and $\phi (1020)$. The parameters
of  $\omega (783)$ and $\phi (1020)$ and their errors  were taken from PDG 
Tables \cite{PDG}. The phases  $\omega (783)$ and $\phi (1020)$
were fixed: $\phi_{\omega (783)}=0;\ \phi_{\phi (1020)}=\pi $.

To evaluate parameters of the $\omega (1600)$-resonance independently
the $e^+e^-\to\omega\pi\pi$ production cross section, measured by DM2
\cite{DM2}, was fitted separately with expression similar to
Eq.\ref{fit}, giving
$M(\omega (1600))=1643\pm 14$~MeV,  $\Gamma(\omega (1600))=272\pm 29$~MeV
and $\sigma_{max}=3.1\pm 0.3$~nb. These values statistically
agree with  the PDG Tables parameters  of $\omega (1600)$,
therefore they  were used in final fitting.

\begin{figure}[htb]
  \epsfig{figure=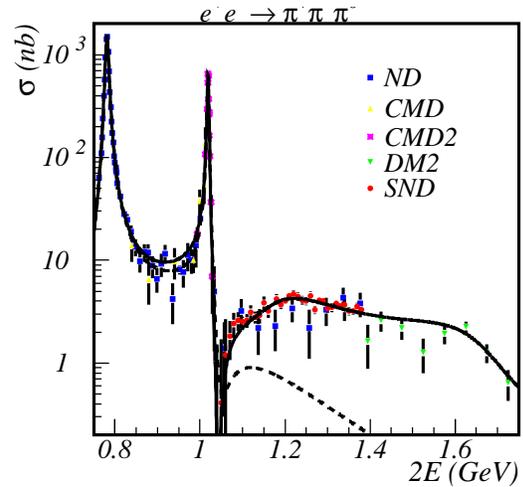,width=0.45\textwidth}
  \caption{\label{p3fit1} The cross section of the process
  $e^+e^-\to\pi^+\pi^-\pi^0$,  measured in different experiments in wide 
  energy range. The solid curve  corresponds to the best fit. The dashed 
  line is a VMD model prediction.}
  \end{figure}

  Four possible choices of  phases $\phi_{\omega (1200)}$ and
$\phi_{\omega (1600)}$:  $\phi_{\omega_i}=0,\pi$, corresponding to
constructive and destructive   interference,  were considered.
We found that the fit with $\phi_{\omega (1200)}=0$ contradicts
CMD-2 data, while the fit with equal $\phi_{\omega (1200)}$ and
$\phi_{\omega (1600)}$ phases disagrees with DM2 data.  The $\omega
(1600)\to 3\pi$ decay in the latter case is not seen. The fit
with $\phi_{\omega_(1200)}=\pi$ and $\phi_{\omega_(1600)}=0$,
satisfies all data.
The data and  resulting fitting curve for this case are shown in
Figs.\ref{p3},\ref{sdm2},\ref{p3fit1}.  The $\chi^2/N.D.$ parameter
for SND data  is 25/34 ($P(\chi^2)$ = 87\%). The main result of the fitting 
is that a new state, referred to as $\omega (1200)$,  was found  instead of
$\omega (1420)$.  The fit  parameters of $\omega (1200)$ and  $\omega (1600)$ 
states are  listed in the Table \ref{tabl2}.
The systematic error $\sim 12$\% is not included into mass and width, but it
is included  into the cross section $\sigma_{max}$ and 
electron width.

    Fitting with other three phase choices, giving poor values of
$P(\chi^2)\sim 5$\%, yields $\omega (1200)$ mass $M_{eff}$, varying within
$1170\div 1250$ MeV and the width $\Gamma_{eff}$ from 190 to 550 MeV.
The optimum interference phases cannot be derived from only SND data,
requiring additional data from outside the interval $2E=1.04-1.38$ GeV
obtained in other experiments.  Thus the interference
phases and $\omega (1200)$  parameters may change if these data
change. Moreover in models with strong energy
dependence of resonance widths the $\omega (1200)$ parameters
are expected to be different too.

    Fitting with a single resonance above  $\phi (1020)$ gives nothing
new, because this case coincides with already mentioned fit with equal
phases $\phi_{\omega
(1200)}$ = $\phi_{\omega (1600)}$ giving poor $P(\chi^2)$. 

Using the  cross section in the $\omega (1600)$ maximum
we obtained the following ratio:

  $ B(\omega (1600)\to 3\pi )/B(\omega (1600)\to\omega\pi\pi )=
   0.17\pm 0.05.$

{\large\bf Discussion.} 
To fit the data above  $\phi (1020)$ we used simple
Breit-Wigner model, which gives resonance parameters (mass, width,..)
close to the visible ones. This approach facilitates the comparison of
data from different experiments. But we keep in mind, that in other models,
e.g. in models with strong width dependence on energy, the mass and width
can differ significantly ($>$ 100 MeV) from their apparent values.    

   If  our explanation  of the cross section enhancement as a new
$\omega (1200)$ state is confirmed,
the question of its nature   arises. It could be either first  radial
excitation $2^3S_1$ or an orbital excitation (D-wave)  $1^3D_1$ of  $\omega
(783)$.  Such excited states are known for $\rho (770)$:  $\rho (1450)$ and
$\rho (1600)$. But here we
encounter the problem of $\omega (1200)$ mass, which is considerably
lower than masses  of its isovector partners. It is
worth mentioning, that the  $\rho (1450)$ and  $\rho (1600)$ parameters are not
well established either. So, new data as well as more advanced analysis
of the existing data are needed. In the energy range
$2E_0<1.4$~GeV in the nearest future  new data are expected from two VEPP-2M
detectors --- SND\cite{SND}  and CMD-2\cite{CMD2} for many channels of
$e^+e^-$ annihilation into hadrons: $2\pi$, $3\pi$, $4\pi$, $5\pi$,
$K\bar{K}$, etc.

{\large\bf Conclusions.} The cross section of the process
$e^+e^-\to\pi^+\pi^-\pi^0$ was measured at VEPP-2M collider by SND detector
in the energy range $2E_0$=1.04 -- 1.38~GeV. The value of the cross section
$\sim$ 4~nb is in good agreement with previous measurements,  but statistical
accuracy is greatly improved. A broad peak  with the mass
$M_{vis}=1220\pm 20$ MeV, referred to as  $\omega (1200)$ is seen.  
The fitting of the cross  section by a sum Breit-Wigner resonances 
with the width not depending on energy, gives  $\omega (1200)$ parameters, 
strongly depending on the interference phases choice.  The PDG  resonance  
$\omega (1420)$ is not  seen in our fitting.

{\large\bf Acknowledgement.} 
    The authors express their gratitude for fruitful
discussions to N.N.Achasov and A.E.Bondar.

  This work is supported in part by STP `Integration', grant No.274 
and Russian Fund for Basic Researches, grants No. 96-15-96327, 
97-02-18563 and 99-02-17155.

\pagebreak
\begin{center}
{\bf Figure captions}
\end{center}
\begin{list}{\textbullet}{}
\item {\it Figure} \ref{vis}:  The visible cross section of the process
    $ e^+e^-\to\pi^+\pi^-\pi^0$. The solid line shows the prediction
       of vector meson dominance model (VMD)
          for the processes (\ref{3pi}) and  (\ref{3pi,gamma}).
\item {\it Figure} \ref{eff}: The  detection efficiency  of the process   (\ref{3pi,gamma}) 
versus radiative photon energy $\omega$.  At the low photon energy  $\omega <$10 MeV
the value of the detection efficiency is the same as for the process  (\ref{3pi}). 
\item {\it Figure} \ref{p3}:
   The $e^+e^-\to\pi^+\pi^-\pi^0$ total cross section,
   measured by VEPP-2M detectors ND \cite{ND}, CMD-2 \cite{CMDP3},
      and SND (this work).  
The only   CMD-2 point at 2E=1040 MeV with the cross section of 0.1 nb 
is not clearly seen,  because it  overlapes  with the nearest SND point.  
The upper solid line is the result of fitting   all existing experimental data according to Eq.\ref{fit}.
		   The lower curve is a VMD model prediction.
\item {\it Figure} \ref{sdm2}: The cross section of the process
     $e^+e^-\to\pi^+\pi^-\pi^0$  from this work  and  DM2 \cite{DM2}
         experiment. The upper solid line is the  fit curve, obtained with
	     Eq.\ref{fit}.  The lower curve is a VMD model prediction.
\item {\it Figure} \ref{p3fit1}: The cross section of the process
  $e^+e^-\to\pi^+\pi^-\pi^0$,  measured in different experiments in wide
      energy range. The solid curve  corresponds to the best fit. The dashed
	    line is a VMD model prediction.
\end{list}

\pagebreak

\clearpage
\pagebreak
\newpage

  \begin{table}
  \caption{\label{tabl1} The cross section   $\sigma_0$ of
  $e^+e^-\to\pi^+\pi^-\pi^0$ process, measured in this work.}
  \begin{tabular}{|l|c|c|r|}
  \hline
  $2E_0$, & $\sigma_0$,nb & $2E_0$, & $\sigma_0$,nb \\
  MeV & & MeV & \\ \hline
  1040 & 0.2$\pm1.3$ & 1050 & 0.4$\pm1.3$ \\ \hline
  1060 & 1.2$\pm1.5$ & 1070 & 1.8$\pm0.8$ \\ \hline
  1080 & 2.5$\pm0.7$ & 1090 & 2.6$\pm0.6$ \\ \hline
  1100 & 2.5$\pm0.3$ & 1110 & 2.6$\pm0.4$ \\ \hline
  1120 & 3.1$\pm0.5$ & 1130 & 2.8$\pm0.4$ \\ \hline
  1140 & 2.9$\pm0.5$ & 1150 & 3.8$\pm0.6$ \\ \hline
  1160 & 3.2$\pm0.4$ & 1180 & 4.2$\pm0.5$ \\ \hline
      1190 & 3.7$\pm0.4$ & 1200 & 4.0$\pm0.3$ \\ \hline
         1210 & 4.5$\pm0.4$ & 1220 & 4.7$\pm0.5$ \\ \hline
	    1230 & 4.5$\pm0.5$ & 1240 & 4.1$\pm0.4$ \\ \hline
	       1250 & 4.0$\pm0.4$ & 1260 & 4.5$\pm0.4$ \\ \hline
	          1270 & 3.3$\pm0.3$ & 1280 & 4.1$\pm0.3$ \\ \hline
		     1290 & 4.0$\pm0.3$ & 1300 & 3.4$\pm0.3$ \\ \hline
		        1310 & 3.5$\pm0.3$ & 1320 & 3.8$\pm0.4$ \\ \hline
		       1330 & 3.6$\pm0.3$ & 1340 & 3.7$\pm0.3$ \\ \hline
		     1350 & 3.2$\pm0.3$ & 1360 & 3.2$\pm0.3$ \\ \hline
		    1370 & 3.5$\pm0.4$ & 1380 & 3.3$\pm0.2$ \\ \hline
		   \end{tabular}
		  \end{table}

   \begin{table}
   \caption{\label{tabl2} The parameters of high mass $\omega$-states, found
   in the fit, described by Eq.\ref{fit} with phase option 
   $\phi_{\omega_(1200)}=\pi$ and $\phi_{\omega_(1600)}=0$}
   \begin{tabular}{|l|c|r|}
   \hline
   Parameter& $\omega (1200)$ & $\omega (1600)$ \\
   \hline
  $M_{eff}$, MeV& $1170\pm 10$ & $1643\pm 14$ \\ \hline
  $\Gamma_{eff}$, MeV&  $187\pm 15$ &  $272\pm 29$ \\ \hline
  $\sigma_{max}$,nb &  $7.8\pm 0.2$  &  $0.54\pm 0.13$ \\
  & $\pm 1.0(syst.)$ &    \\   \hline
  $\Gamma_{\omega ee}\cdot B_{\omega
  3\pi }$ , eV & $137\pm 3$ & $27\pm 7$ \\
  & $\pm 15(syst.)$ &   \\  \hline
  \end{tabular}
  \end{table}

\end{document}